# FRACTALITY AND SELF-ORGANIZATION IN THE ORTHODOX ICONOGRAPHY

Miloš Milovanović
Mathematical Institute of the Serbian Academy of Sciences and Arts,
Knez Mihailova 36, 11000 Belgrade, Serbia
milosm@mi.sanu.ac.rs

Bojan M. Tomić
Institute for Multidisciplinary Research, University of Belgrade,
Kneza Višeslava 1, 11030 Belgrade, Serbia
bojantomic@imsi.rs

**Keywords:** self-organized systems, fractal geometry, self-similarity, wavelets, hidden Markov model, iconography

**Abstract**
The authors consider the Orthodox iconography of Byzantine style aimed at examining the existence of complex behavior and fractal patterns. It has been demonstrated that fractality in icons is manifested as two types – descending and ascending, where the former one corresponds to the apparent information, and the latter one to the hidden causal information defining spatiality of the icon. Self-organization, recognized as the increase of the causal information in the temporal domain, corresponds to contextualization of the observer`s personage position. The results presented in the forms of plots and tables confirm the adequacy of the model being the completion of visual perception.

1. **Introduction**

In this journal many articles dealing with the existence of fractal patterns in various natural and social phenomena have been published [1-4]. The intent of the paper is to complete the series considering the Orthodox iconography of Byzantine style that is the synthesis and the ultimate reach in the medieval tradition of the Eastern Roman Empire for a period of over thousand years, but also of the entire Christian East till nowadays. Though not regarded in this sense, it can be considered that the topic appears not as an ephemeral, but a paradigmatic issue since it does not correspond to a cultural or social phenomenon only, but to a deliberate theological message that can be elucidated just in the context of complex behavior and fractal geometry.

The basic definition of self-organization originates from Shalizi [5] and uses the concept of statistical complexity introduced by Grassberger [6]. Milovanović et al. have developed an effective algorithm for quantifying complexity of large signals based on decomposing of the signal using wavelet transform [7]. Applied to artworks, it has been demonstrated that self-organization regularly occurs and corresponds to the creative process in art [8]. Finally, in considering the iconography the aim is to demonstrate not only that self-organization occurs, as well as in other artworks, but also that one deals with *personalized geometry* revealing the observer's personality as the origin of fractal geometry of icon. In other words, the complex behavior corresponds to contextualization of the observer`s personage position rising to the concept of personality which is the essential moment of an icon and to which the entire geometrical structure is subjected.

The matter is exposed in the following way. In Section 1 the subject and method of the study are presented, and also the order of exposing the matter. In Section 2 some examples of self-similarity in icons are given with establishing the concept of *personalized geometry*. Section 3 considers self-organization in icons aimed at establishing an adequate conceptual framework for its interpretation which is the topic of Section 4. Section 5 contains plots of self-organization as the increase of complexity in temporal domain for all of icons mentioned in the paper. In Section 6 there are some concluding remarks.

## 2. Examples of Self-Similarity in Icons

Under the term fractality in this paper the authors consider self-similarity, i.e., the emergence of the same figure at different scales, being one of defining features of the term. Although originally geometric one, the definition can be conceived in an extended sense including biological structures, as well as diverse social and cosmic ones. It can be generally comprehended as a generative property defining in each particular case the specific fractal geometry of a considered matter.

The review of fractality in icons commences with The Presentation of the Virgin (Fig. 1). One observes that the figure of the Virgin emerges twice – both on the bottom and on the top of the staircase. The existence of the staircase induces an extra depth dimension of the image that indicates scaling, but also establishing a time axis since the icon theme is the entrance of the Virgin into the temple which is the event of her ascending up the stairs to the throne. The fractality is also present in the fact that a three-year-old girl is represented as a reduced figure of the adult person, i.e., as the Virgin. Consequently, it is about a specific geometry where the emergence of the same figure at different scales indicates the personality dynamics. This is referred to as personalized geometry of icon with such reservation that the complete meaning of the concept will be clarified further on as it depends on the other elements of the paper.

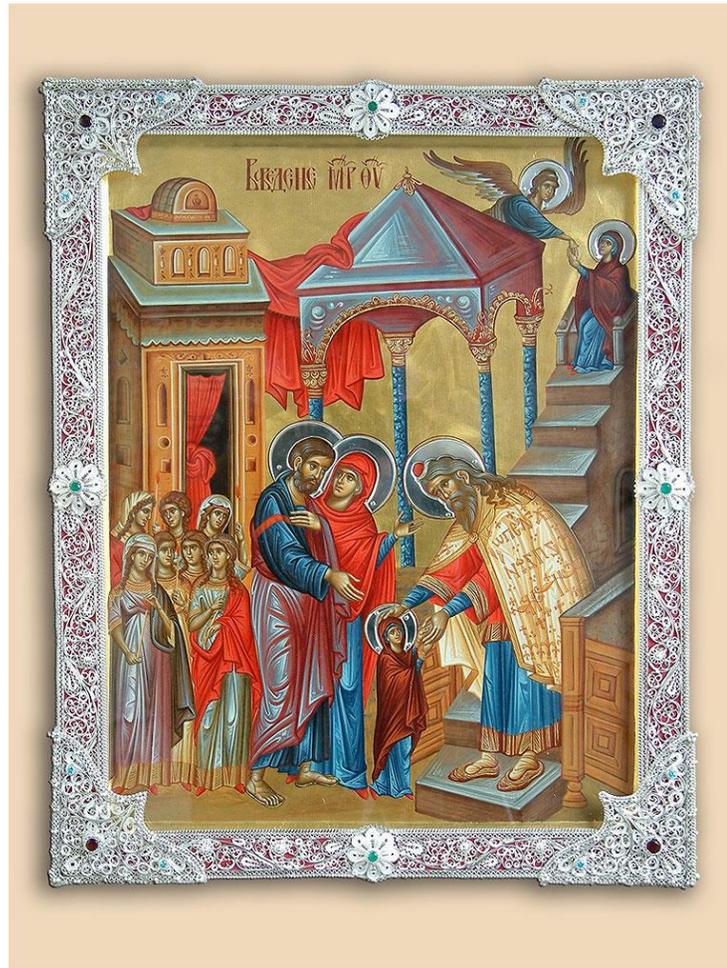

Figure 1. The Presentation of the Virgin. The figure of the Virgin emerges twice – both at the bottom and the top of the staircase that indicates scaling but also establishing time axis

Another example for the matter concerns the icon of the Dormition of the Virgin (Fig. 2) where the figure of the Virgin emerges at two scales – lying on the catafalque and on the hands of Christ. Some variants of the same icon contain three scales (Fig. 3), i.e., the Virgin emerges on the catafalque in a horizontal position, then on the hands of Christ and finally in a vertical position ascended to heaven.

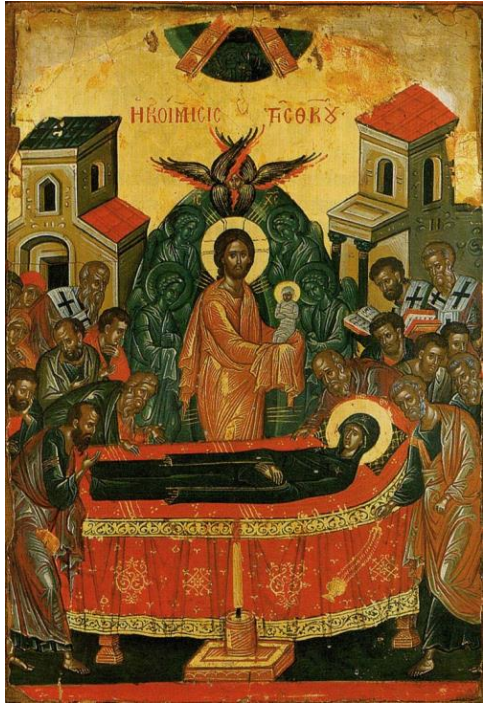

Figure 2. The Dormition of the Virgin. Figure of the Virgin emerges at two scales – lying on the catafalque and on the hands of Christ

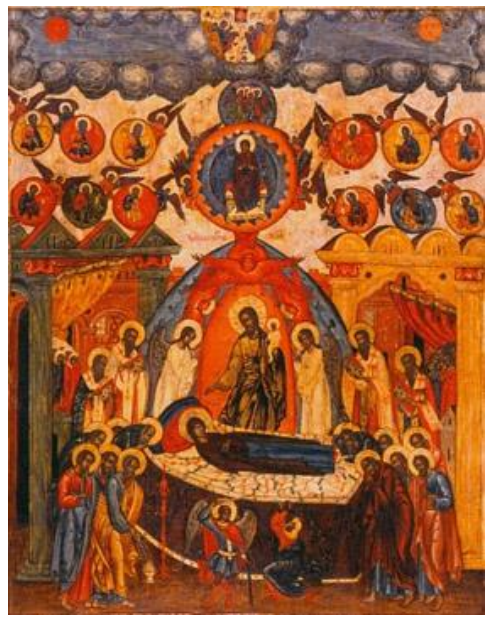

Figure 3. The Dormition of the Virgin. The figure of the Virgin emerges at three scales – on the catafalque, then on the hands of Christ and finally ascended to heaven

The next example concerns the icon of John the Baptist (Fig. 4) where his figure emerges at two scales – at the bottom of the image beheaded and also as alive and erected. There is a variant containing three scales (Fig. 5) and once again the erection from the horizontal to the vertical position is observed.

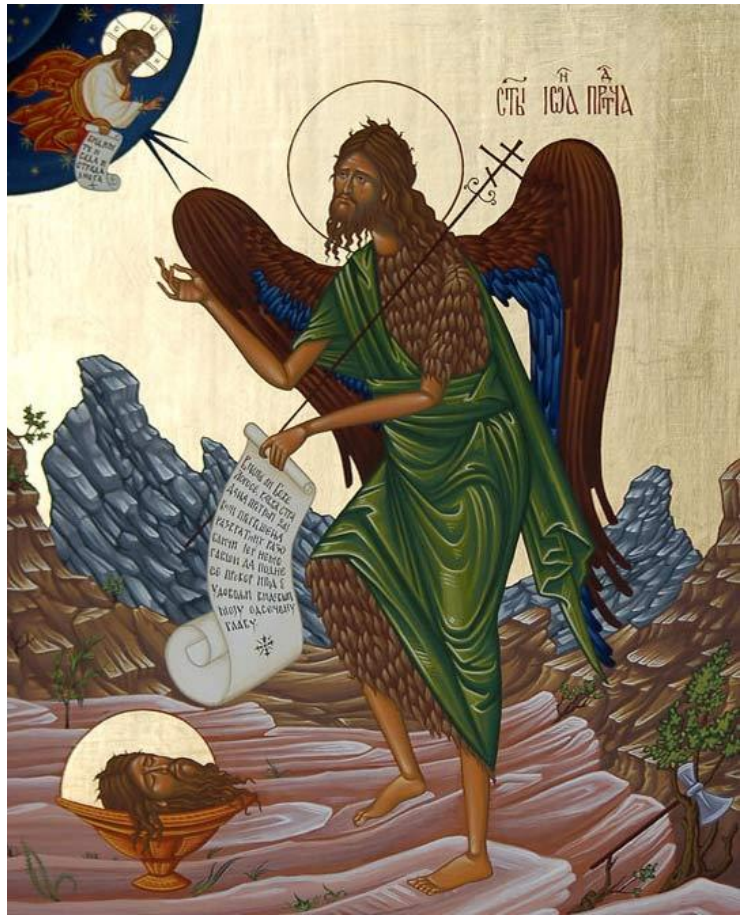

Figure 4. The Beheading of John the Baptist. The figure emerges at two scales – at the bottom as beheaded, and also as alive and erected

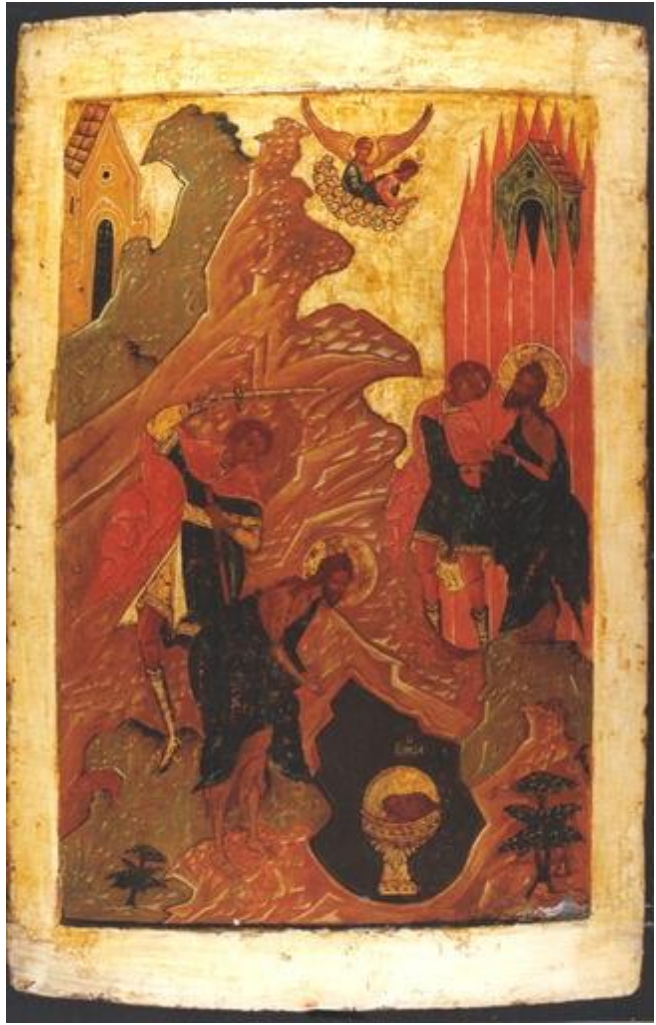

Figure 5. The Beheading of John the Baptist. The figure emerges at three scales and the erection from the horizontal to the vertical position is observed

In the icon of the Transfiguration (Fig. 6) one observes von Koch`s curve that indicates scaling in which Christ`s personality is revealed to the apostles as the Uncreated Tabor Light. One should note that at the same scale with the apostles in the left lower corner also emerges a relief element conveying fractality to the terrestrial relief presented in the icon through fractured and cracked rocks.

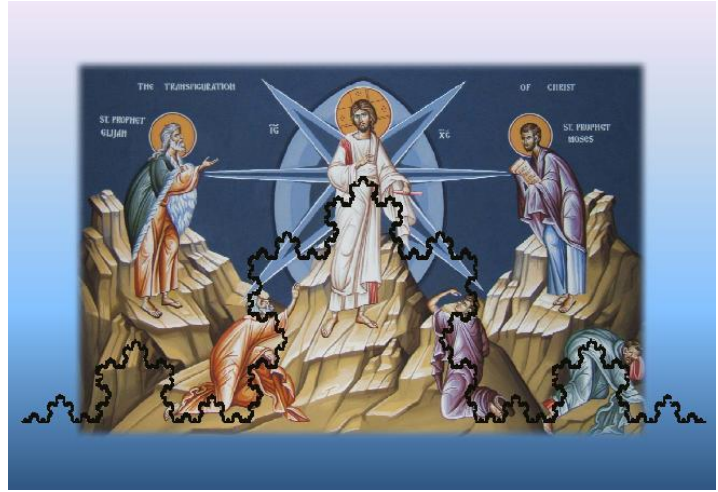

Figure 6. The Transfiguration. One observes von Koch`s curve that indicates scaling in which the Christ`s personality is revealed to apostles as the Uncreated Tabor Light

Finally, the last example in the section concerns the icon of The Last Supper (Fig. 7). The self-similarity is indicated in the emergence of Christ both at the table with the apostles and on the table in the form of the Holy Grail representing his blood and body. The icon actually contains one more scale since it is placed by default in the iconostasis in front of which the Communion takes place and thus the Holy Grail emerges once again in the actual rite context of the followers` Communion. Thereby at the last scale of the icon occurs the observer as communicant of divine personality disclosing fractality as a principle of personality existence, that unifies the divine and the human (Fig. 8) through the formula personality-communicant-observer.

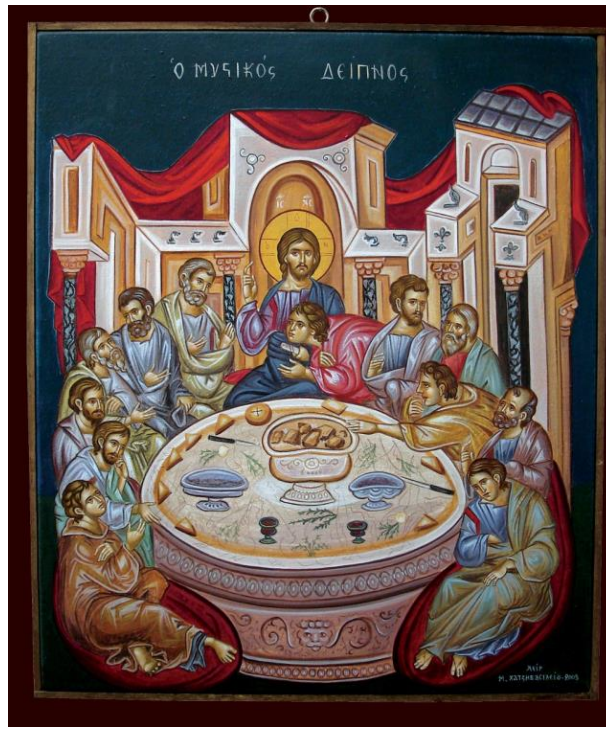

Figure 7. The Last Supper. Christ emerges both at the table and on the table in the form of the Holy Grail

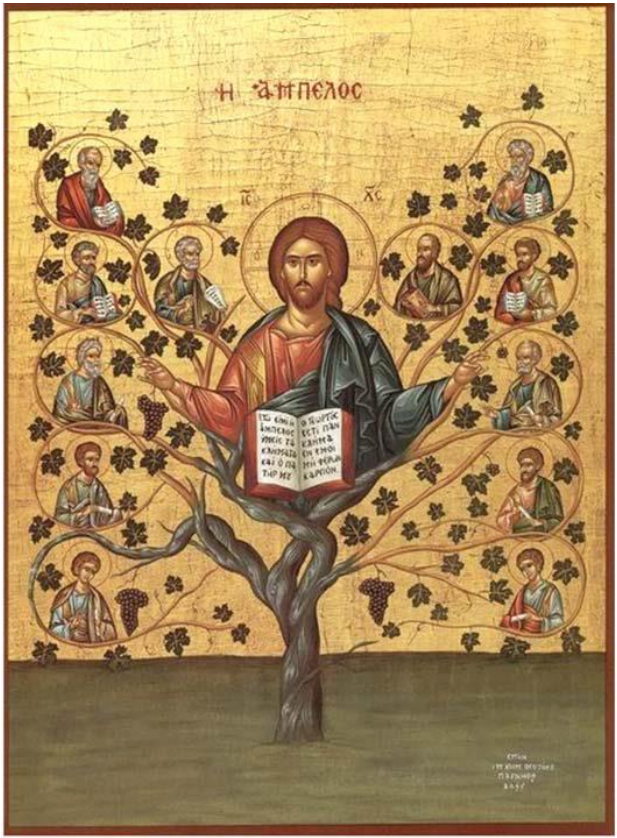 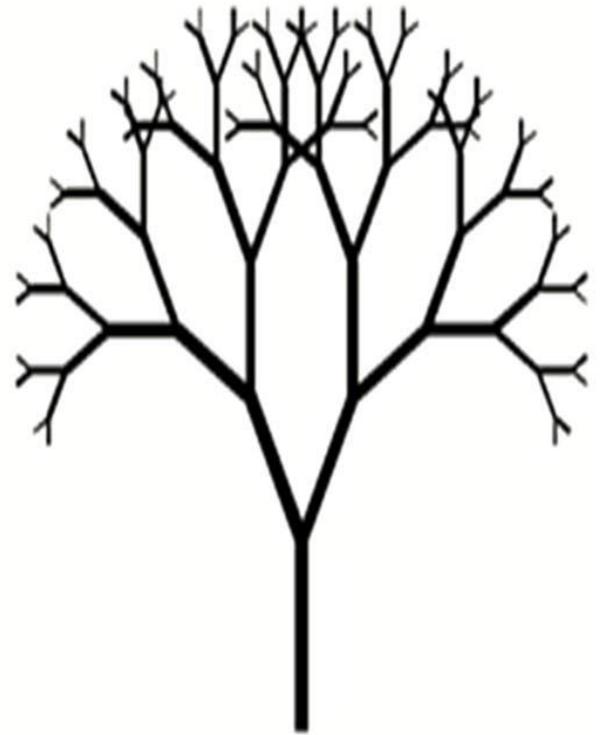

Figure 8. The Tree of Life unifying human personalities with divine one. Each figure of the icon is an icon for itself disclosing the fractal tree structure

### 3. Time and Self-Organization in Icons

The matter of time in panting has been fundamentally discussed in Souriau's article "Time in the Plastic Arts". At first he indicates that nothing is more dangerous for the exact and delicate understanding of the plastic arts (design, painting, sculpture, architecture) than their rather banal description as "arts of space" in contrast to the phonetic and cinematic arts (music, poetry, dance, film) characterized as "arts of time" [9]. The widely accepted distinction popularized by Lessing, is considered to be flawed, since it is unable to account for the complex nature of the arts [10]. In order to overcome the evident deficiency of his approach Souriau introduces the concept of *intrinsic time* defining it as "artistic time inherent in the texture itself of a picture in its composition, in its aesthetic arrangement" [9]. Moving to the examples and illustrations he uses fail to show what exactly constitutes "intrinsic time" [11]. Applying his concept to consider spatiality of icon Clemena Antonova claims that time is a fundamental organizing principle of pictorial art [10]. In her noteworthy attempt something essential was missing – an adequate conceptual framework that would make it possible to comprehend the complex nature of icon.

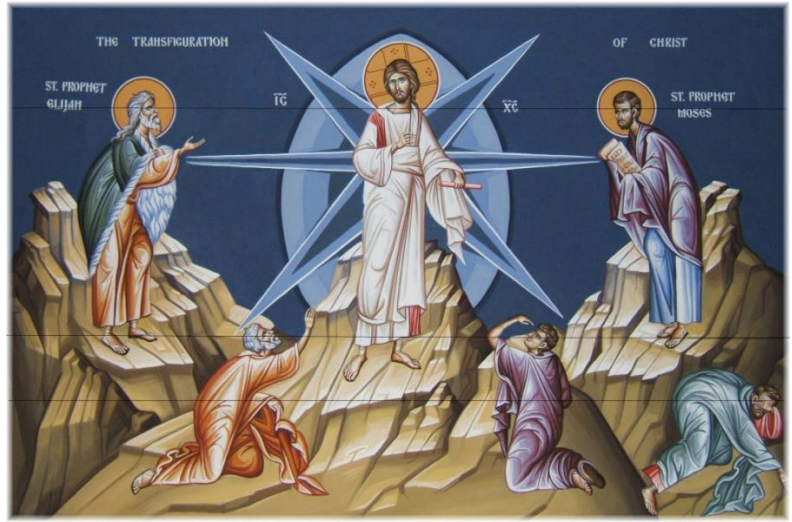

Figure 9. The Transfiguration. The expansion of the mountain massifs establishes scaling where the personality dynamics is recognized in the progression from the horizontal through the semi-vertical to, finally, vertical position in respect to central Christ`s figure

When considering the icon of Presentation (Fig. 1), it was indicated that the *personalized geometry* induces the time axis with scaling property as the fundamental principle of fractal organization of the icon. The adjoint depth dimension is manifested through the icon expansion towards the interior, being a general characteristic of the Byzantine iconography, which, due to missing of a more adequate term, is referred to as the reverse perspective (in Russian обратная перспектива). The term is due to Russian mathematician Pavel Florensky (Па́вел Алекса́ндрович Флоре́нский, 1882 – 1937) who established it in opposition to the modern linear perspective pointing out the image contraction towards the interior [12]. The same concept was introduced almost twenty years afterwards by Serbian mathematician Miloš Radojčić (Милош Радојчић, 1903 – 1975) who designated it counter perspective (in Serbian контраперспектива) [13]. However, that one has nothing to do with the mere reversion of the linear perspective it can be already seen from the example of Fig. 1, where together with the expansion perspective of the icon the contraction perspective of the staircase is observed concurrently. Thus both perspectives coexist in the same image and, consequently, they are not opposed or exclusive mutually.

The concept of time as a scale makes it possible to put the spatiality of icon in the context of self-organization model, which is the next section topic. In this context time appears as the axis of system self-organizing that corresponds to overcoming of causal necessity in the model which correlates the icon geometry with the observer's personality. Actually, the personality is constituted through the overcoming of causal necessity [14], which is the formulation that can be regarded as the definition of personality dynamics. At this point a specific fractality also takes place with self-similarity that implies scaling of the observer's personality progressing along the time axis. In such a way the *personalized geometry* of the icon reveals the observer's personality as its essential moment and its origin.

An example of the mentioned above is the icon of Transfiguration with the reverse perspective depth dimension implied by expansion of the mountain massifs towards the image interior. The time axis recognized in such a way establishes scaling where the personality dynamics is recognized in progression from the horizontal through semi-vertical to finally vertical position with respect to the central Christ`s figure (Fig. 9). One should note that the scaling like this is different from that mentioned above in the consideration of the same icon (Fig. 6). In that respect one discriminates two types of fractality in icon – the *descending* and the *ascending* one. The *descending fractality* concerns the contents of icon revealing the personality of Christ or of the saint represented in the icon. The *ascending fractality*, however, concerns the self-organizing spatiality revealing the observer's personality through the reverse perspective scaling. Using the model language that is going to be introduced in the next section, it can be said that the former fractality type concerns the apparent information, whereas the latter one concerns the hidden causal information. The *descending fractality* may be sidelined, which is not rare in the case of icons with a simple apparent composition. The *ascending fractality* is always present which agrees well with the assertion of Clemena Antonova that the reverse perspective is the defining feature of icon [10, p. 29].

Another example is the icon of The Last Supper. The scaling starts from the observer as an outer space object and going through the Grail and table it ends on Christ's chest where the head of Saint John the Apostle is (Fig. 10). The reverse perspective rays intersect at some relevant elements of the image [15], but they also emerge out from the image establishing a unique spatial framework through formula observer-communicant-personality. Accordingly, the icon space is not limited to the mere content of the icon, but also includes the outer space pointing out the observer as the principal personality of the apparent world in which the icon also takes place objectively.

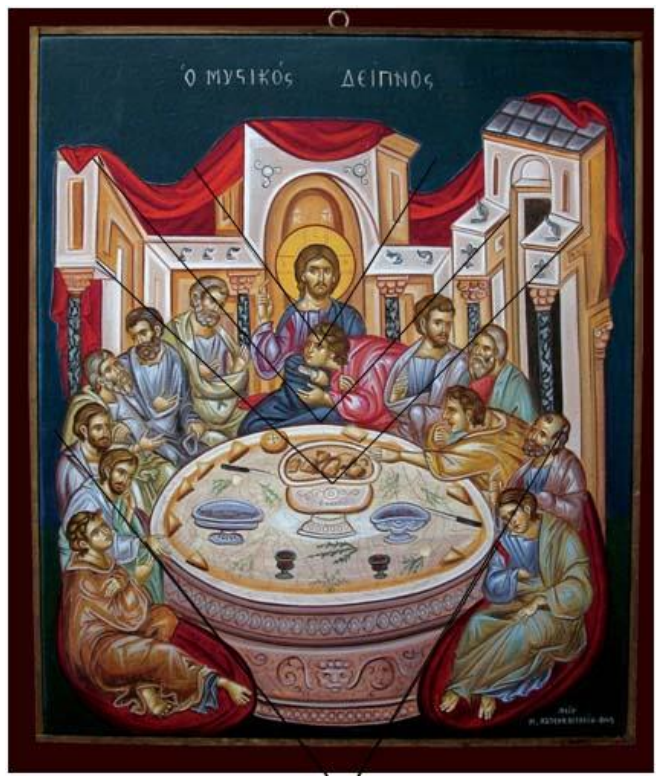

Figure 10. The Last Supper. The scaling starts from the observer as an outer space object and going through the Grail and table it ends on Christ's chest

## 4. Quantifying Self-Organization with Wavelets

The pursuit for the self-organization model of the icon commences from its representation as a two-dimensional signal in grayscale mode. In this way the focus has been placed on the spatial structure and lightening, whereas the colors are marginalized according to the fact that they are without any significant attention in this paper. Following the same line with previous insights it is required an icon to be considered as a spatio-temporal signal with temporal domain corresponding to the scale. This actually means that the signal should be represented in both spatial and frequency domains concurrently, which is realized by applying the wavelet transform that has been already used as a tool in the research of self-organization [7]. The suggested model is, thus, completely developed in the corresponding literature, and it has been demonstrated as widely applicable to various signals. Here it is presented just briefly, at first for the case of one-dimensional signals.

A wavelet transform decomposes a one-dimensional signal $F(x)$ in terms of translated and dilated versions of the band-pass wavelet function $\psi(x)$ and translated versions of the low-pass scaling function $\phi(x)$ [16]. For a signal of a dyadic dimension $J$ (i. e. length $2^J$) the decomposition is

$$F = A \cdot \phi_{[J]} + \sum_{j=-J}^{-1} \sum_{k=0}^{2^{J+j}-1} D_{j,k} \cdot \psi_{j,k},$$

where $j$ indexes dyadic scale, and $k$ indexes the spatial location in the basis elements $\psi_{j,k}(x) = \psi(2^j x - k), \phi_{[J]}(x) = \phi(2^{-J} x)$. For a wavelet $\psi(x)$ centered on a frequency $\xi_0$, the detail coefficient $D_{j,k}$ measures the signal content around place $2^{-j} k$ and frequency $2^j \xi_0$. In this way one obtains a signal representation as an approximate coefficient $A$ and pyramid of detail coefficients $D = (D_i)$, where for simplicity a pseudo-numeration of knots is introduced from the top downwards successively (Fig. 11). Thereat every coefficient on the scale $j < -1$ has two successor coefficients on the next scale $j+1$ sharing its spatial support which structures the pyramid in the form of the binary tree.

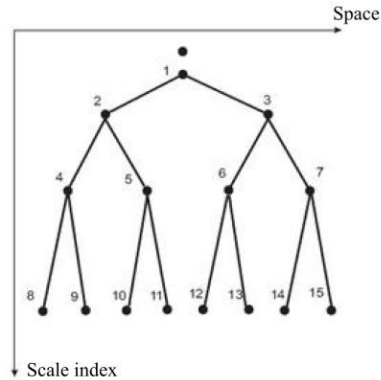

Figure 11. Signal representation in the form of an approximate coefficient and the tree of detail coefficients

The statistical model of the detail coefficients tree is founded on the wavelet transform properties of which the basic one is approximate decorrelation [17]. This means that in such representation the correlations mostly concern successions along the tree branches. Furthermore, the correlations between the tree knots $D = (D_i)$ occur only through the hidden tree variables $S = (S_i)$ whose knots are referred to as local causal states. So a hidden Markov model is formed (Fig. 12) which has been demonstrated as widely applicable in the signal processing [17]. In the basic case two values are enough to be taken by the hidden variable $S_i$ at a particular knot, but the model can be easily generalized to an arbitrary number of hidden states values assuming a more refined causal structure. The model parameters involve the parameters of Markov's hidden states tree, as well as the parameters of conditional coefficient distributions which are assumed to be Gaussian for given hidden state value at a knot.

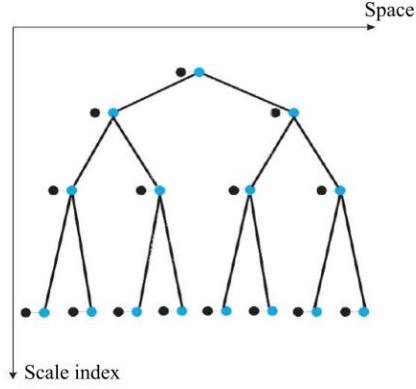

Figure 12. Hidden Markov model for the tree of detail coefficients. The black color is used for the detail coefficients and the blue one for the hidden states

Exhausting all correlations in the signal the causal states contain minimal information required for optimal prediction, being in line with the concept of statistical complexity as defined by Grassberger [6]. This makes it possible to consider the complex behavior in the tree of detail coefficients proving that the global complexity $C = H(S)$ is a measure of self-organization [7] understood as an increase of the local complexity $C_i = H(S_i)$ in temporal domain [5]. The temporal domain corresponds to dyadic scale as the intrinsic time of the model since the correlation succession of the coefficients takes place along the vertical axis of the tree. Under information the authors imply Shannon's entropy $H(\cdot) = -\sum_n p_n \cdot \log p_n$ which is an extensive uncertainty measure for the random variable.

In this sense the complexity as the information content of the causal state appears as its statistical uncertainty opposed to the classical causality understood as necessity with zero informational content. Therefore, the self-organization as complexity increase in temporal domain is in line with overcoming the causal necessity in the model. The complete information of the signal is decomposed according to the canonical equality $H(D) = H(S) + H(D|S)$ where the first term is the measure of the system complexity, whereas the second one means noise since it corresponds to the uncertainty which is present even in the case when all the correlations are recognized [7]. It is important to note that the complexity recognized in such a way depends on the basis in which the process is conceived indicating the observer as the principal personality of the system since the complex behavior depends on observer`s recognition. In other words, self-organization is manifested as a principle of a complex relationship between the personality and the apparent world, whereby the personality appears as no constituent of the complex system, but acts as its completion being manifested through the causal information increase in temporal domain.

The model is applied on two-dimensional signals with minimal modifications that concern the wavelet transform of the two-dimensional signal $F(x,y)$. For a signal of dyadic dimension $J$ (i. e. length $2^J x 2^J$) the decomposition is

$$F = A \cdot \phi_{[J]} + \sum_{j=-J}^{-1} \sum_{k=(0,0)}^{(2^{J+j}-1, 2^{J+j}-1)} (D_{j,k}^h \cdot \psi_{j,k}^h + D_{j,k}^v \cdot \psi_{j,k}^v + D_{j,k}^d \cdot \psi_{j,k}^d)$$

so that in the signal representation, in addition to the approximate coefficient $A$, there are three pyramids of detail coefficients – horizontal $D^h = (D_{j,k}^h)$, vertical $D^v = (D_{j,k}^v)$ and diagonal one $D^d = (D_{j,k}^d)$, and each of them has the quadtree structure and its own complex behavior.

## 5. Results

In this section the results of the model applied to icons mentioned above in the paper are presented. The applied model is with five values of hidden state variables and its parameters for apparent signal values are determined using Baum-Welch algorithm [17] making it possible to calculate the local and global complexities. In the plots the diagonal (upper left), horizontal (upper right), vertical (lower left) complexities are given, as well as their arithmetic mean (lower right). The represents of different wavelet families and the corresponding scaling functions are presented in different colors: Haar (haar) – black, Daubechies (db2) – red, Symlet (sym3) – yellow, Coiflet (coif1) – green, Biorthogonal (bior1.3) – blue, Reverse biorthogonal (rbior1.3) – magenta, Discrete Meyer (dmey) – cyan. The values for global complexities with bold maximal values corresponding to the optimal base are given in the tables. The plots are normalized to a zero mean value and unit variance, whereas in the case of the global complexity the information measure is also normalized to unity.

The particular plots concerning different wavelets are somewhat different, though something common in the behavior is seen which is more obvious for the mean complexity plots (lower right). Generally speaking a significant increase in the local complexity on the first two, three or four scales for each plot can be observed, corresponding to self-organization that appears through reverse perspective expansion of icon. The maximal complexity values in the tables result in a rather strict order among the signals. Its meaning can be understood to some extent if we compare the variants of the same icon with two or three scales (pairs Fig. 14-15 and Fig. 16-17) where it is seen that a larger number of scales noted on an icon corresponds to a higher maximal global complexity in each table. However, one should be aware that self-organization does not correspond to the signal apparent information, but to the hidden causal information defining own spatiality of icon. In this respect the results appear as a completion of the visual subject observation yielding a quantitative description and objective meaning to personality dynamics implied and revealed by the icon.

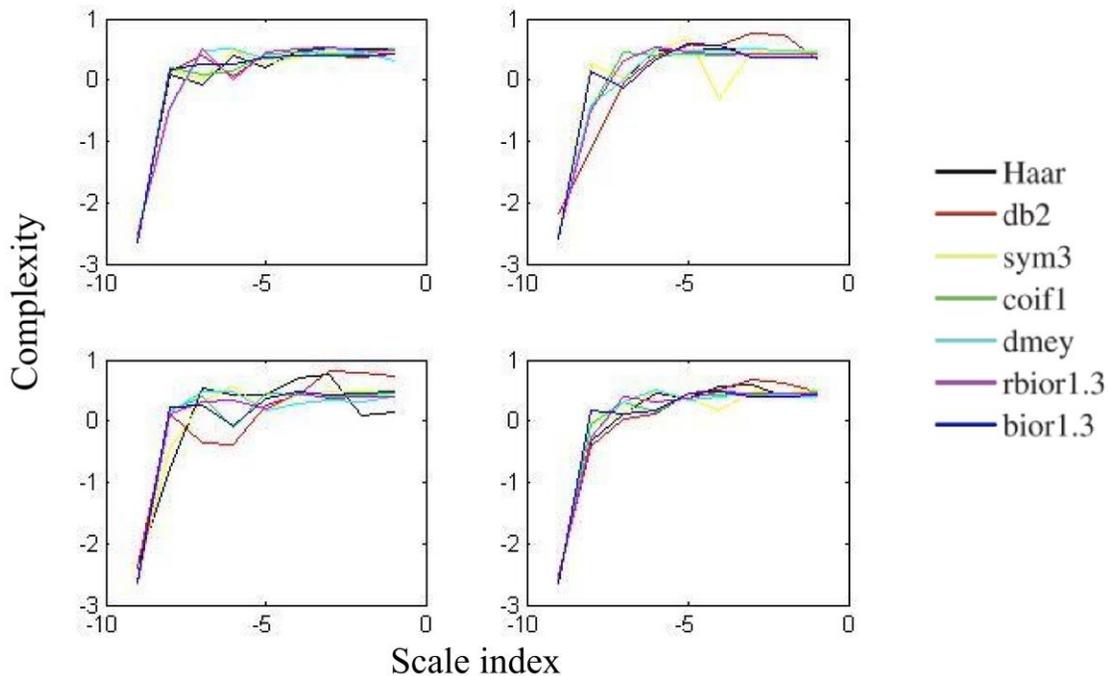

Figure 13. Local complexity plots for The Presentation of the Virgin: diagonal (upper left), horizontal (upper right), vertical (lower left) and mean (lower right)

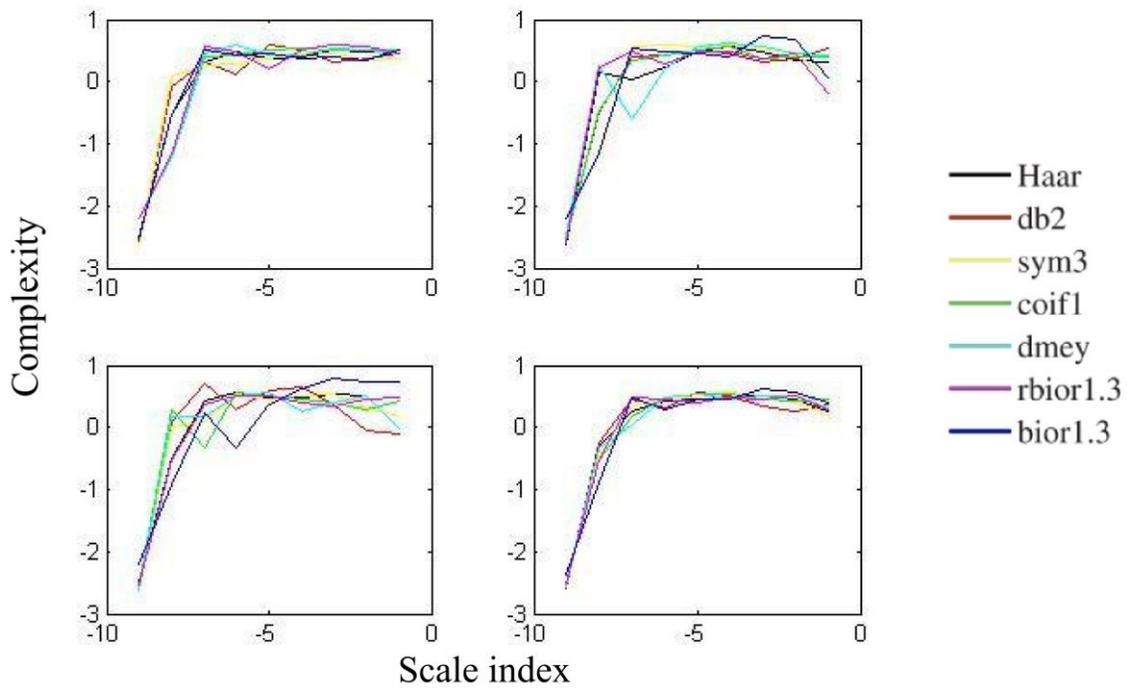

Figure14. Local complexity plots for The Dormition of the Virgin (first one): diagonal (upper left), horizontal (upper right), vertical (lower left) and mean (lower right)

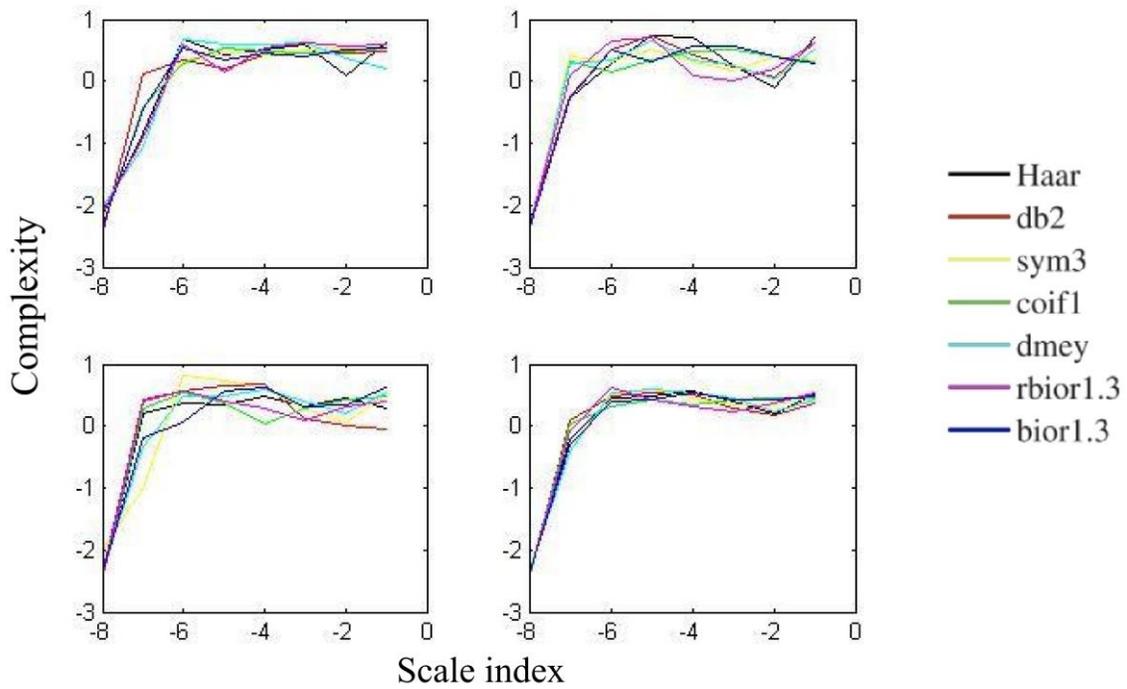

Figure 15. Local complexity plots for The Dormition of the Virgin (second one): diagonal (upper left), horizontal (upper right), vertical (lower left) and mean (lower right).

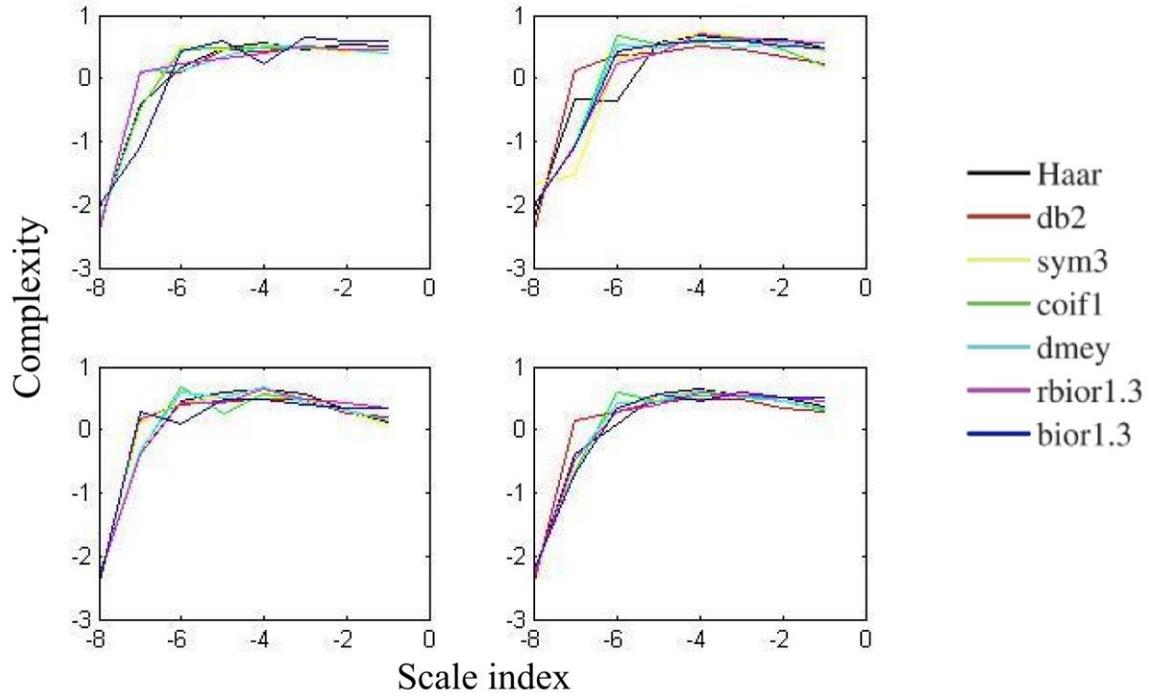

Figure 16. Local complexity plots for The Beheading of John the Baptist (first one): diagonal (upper left), horizontal (upper right), vertical (lower left) and mean (lower right).

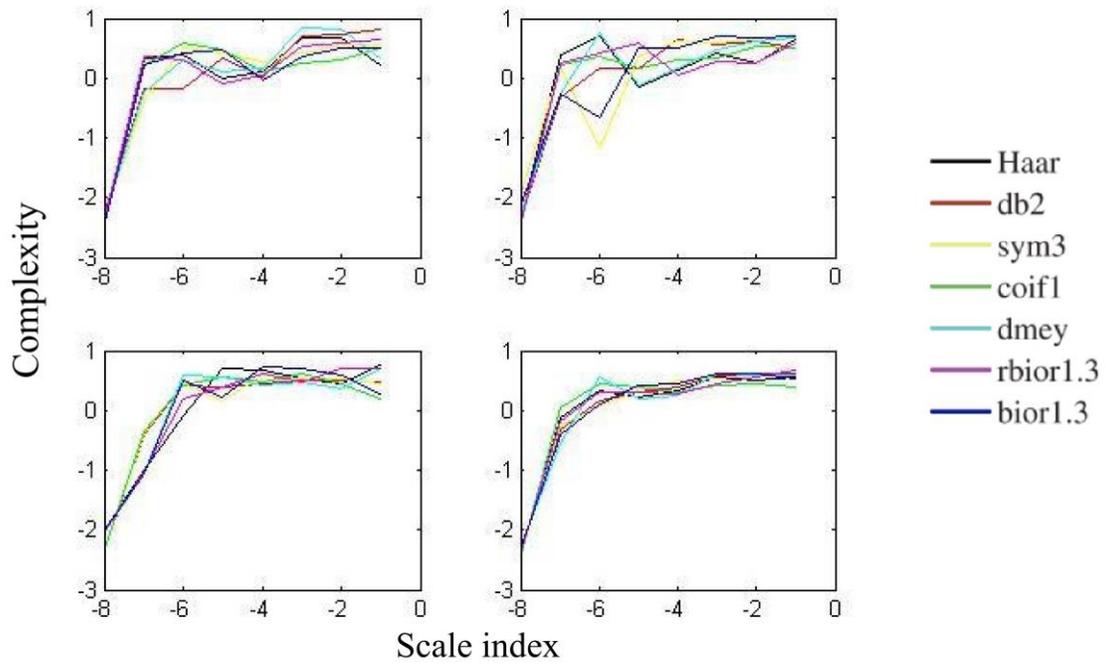

Figure 17. Local complexity plots for The Beheading of John the Baptist (second one): diagonal (upper left), horizontal (upper right), vertical (lower left) and mean (lower right)

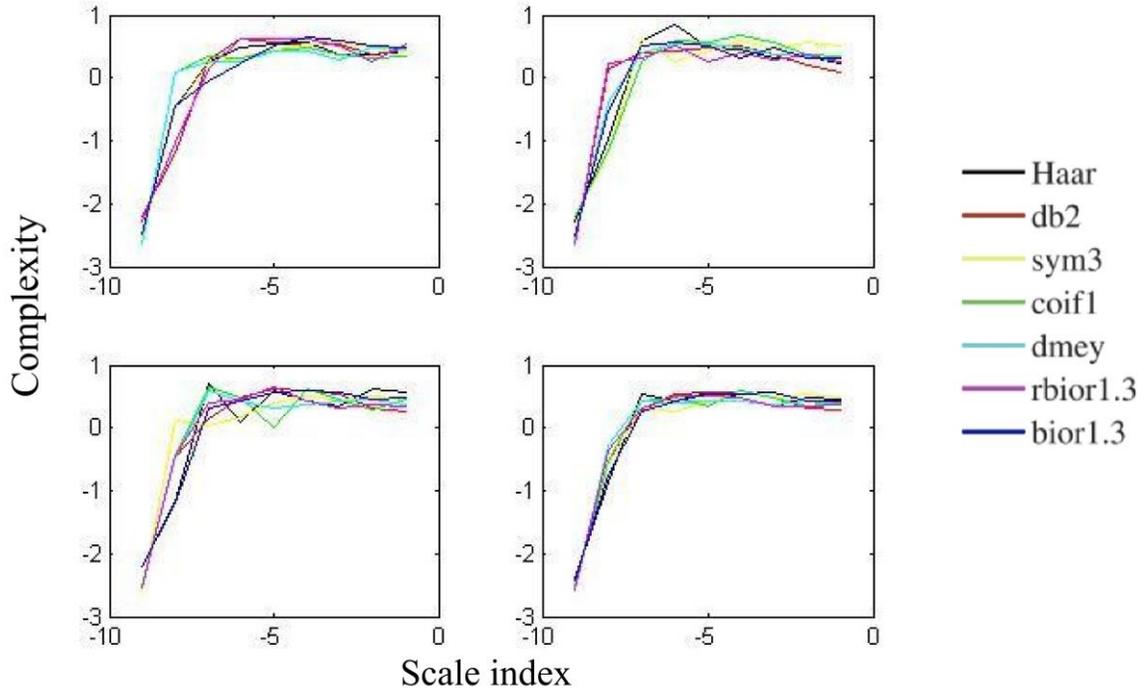

Figure 18. Local complexity plots for The Transfiguration: diagonal (upper left), horizontal (upper right), vertical (lower left) and mean (lower right)

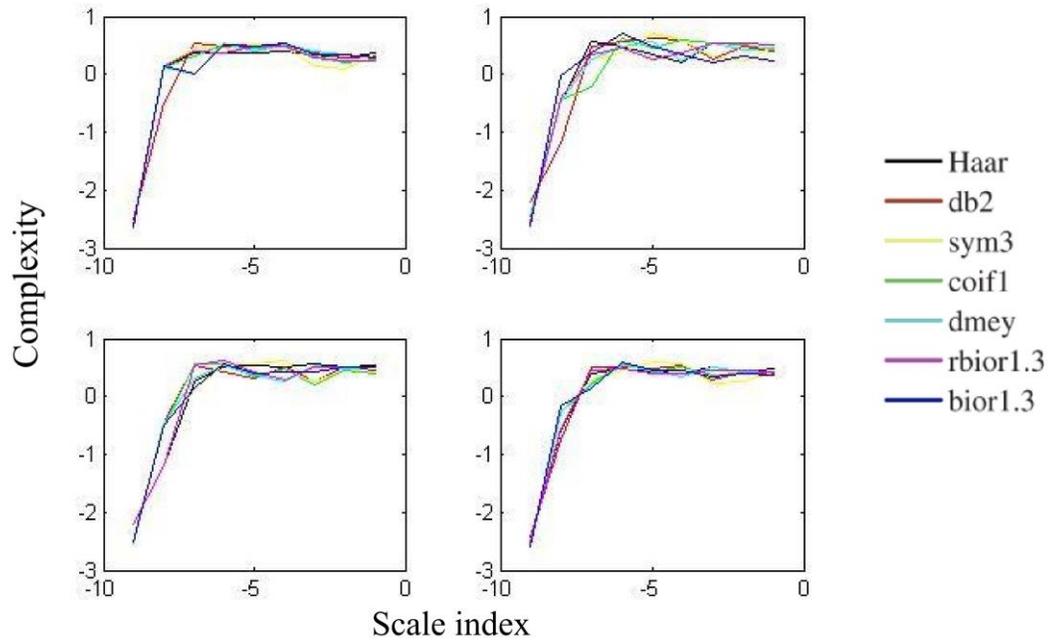

Figure 19. Local complexity plots for The Last Supper: diagonal (upper left), horizontal (upper right), vertical (lower left) and mean (lower right)

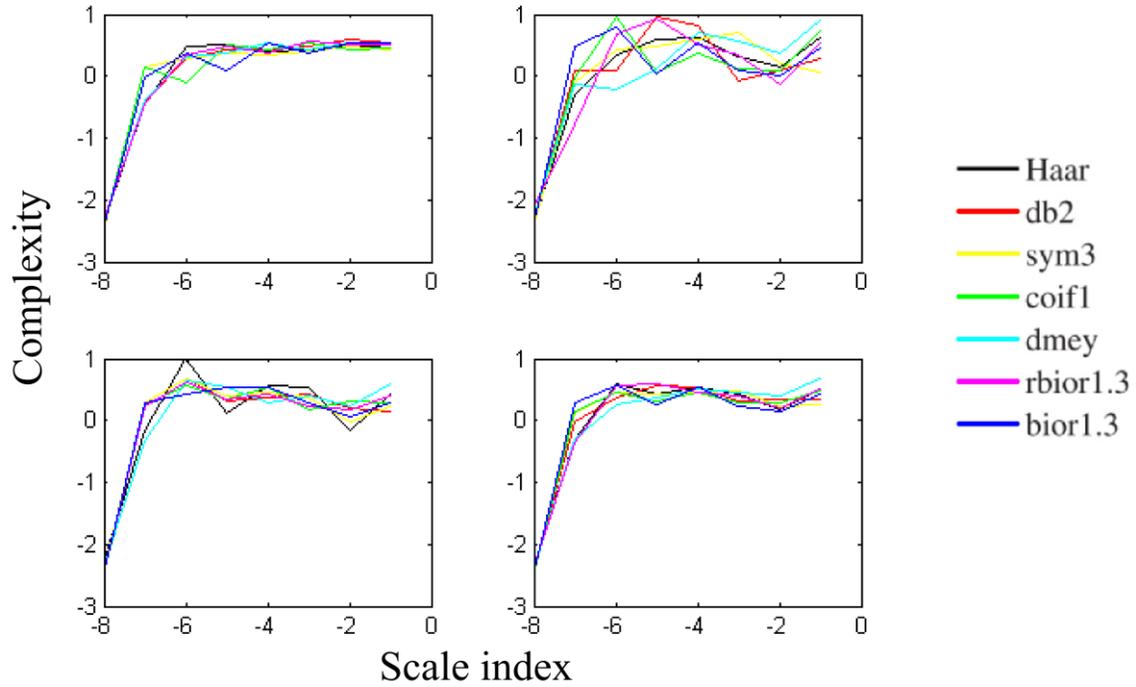

Figure 20. Local complexity plots for The Tree of Life: diagonal (upper left), horizontal (upper right), vertical (lower left) and mean (lower right)

*Tables*

Table 1. Values of diagonal global complexity for corresponding plots

|         | Fig. 13 | Fig. 14 | Fig. 15 | Fig. 16 | Fig. 17 | Fig. 18 | Fig. 19 | Fig. 20 |
|---------|---------|---------|---------|---------|---------|---------|---------|---------|
| Haar    | 0.2009  | 0.2997  | 0.2355  | 0.2178  | 0.1335  | 0.3279  | 0.2303  | 0.3465  |
| Db2     | 0.2801  | 0.4380  | 0.4895  | 0.3239  | 0.4753  | 0.3322  | 0.4298  | 0.4976  |
| Sym3    | **0.3224** | **0.5153** | **0.5580** | **0.3631** | **0.5348** | 0.3162  | **0.4779** | **0.5138** |
| Coif1   | 0.2690  | 0.4712  | 0.4301  | 0.2782  | 0.3897  | **0.4163** | 0.4070  | 0.4168  |
| Bior1.3 | 0.1917  | 0.2259  | 0.1522  | 0.1879  | 0.1288  | 0.2592  | 0.2462  | 0.2882  |
| Rbio1.3 | 0.2209  | 0.3540  | 0.3351  | 0.2061  | 0.3159  | 0.3343  | 0.3131  | 0.3166  |
| Dmey    | 0.2066  | 0.5071  | 0.4660  | 0.1960  | 0.3398  | 0.4030  | 0.3362  | 0.3726  |

Table 2. Values of horizontal global complexity for corresponding plots

|         | Fig. 13 | Fig. 14 | Fig. 15 | Fig. 16 | Fig. 17 | Fig. 18 | Fig. 19 | Fig. 20 |
|---------|---------|---------|---------|---------|---------|---------|---------|---------|
| Haar    | 0.2289  | 0.2270  | 0.4252  | 0.3390  | 0.3909  | 0.2516  | 0.2041  | 0.4546  |
| Db2     | 0.2607  | **0.3290** | 0.3458  | 0.2962  | 0.3237  | 0.3053  | 0.2743  | 0.2804  |
| Sym3    | **0.3089** | 0.2013  | 0.2692  | **0.3418** | **0.4222** | **0.3689** | **0.3967** | 0.2690  |
| Coif1   | 0.2754  | 0.1803  | 0.2453  | 0.2888  | 0.2695  | 0.3274  | 0.2456  | 0.3296  |
| Bior1.3 | 0.2448  | 0.2025  | **0.4642** | 0.3414  | 0.4050  | 0.2308  | 0.1890  | **0.4634** |
| Rbio1.3 | 0.2141  | 0.1341  | 0.4432  | 0.2862  | 0.3268  | 0.2522  | 0.1946  | 0.3356  |
| Dmey    | 0.1662  | 0.1446  | 0.2642  | 0.1690  | 0.2711  | 0.3108  | 0.1836  | 0.2374  |

Table 3. Values of vertical global complexity for corresponding plots

|         | Fig. 13 | Fig. 14 | Fig. 15 | Fig. 16 | Fig. 17 | Fig. 18 | Fig. 19 | Fig. 20 |
|---------|---------|---------|---------|---------|---------|---------|---------|---------|
| Haar    | 0.2292  | 0.1918  | 0.3971  | 0.2670  | 0.3301  | 0.2687  | 0.2187  | 0.3850  |
| Db2     | **0.3474** | 0.2072 | 0.1394 | 0.2596 | 0.3208 | 0.2986 | 0.2694 | 0.3300 |
| Sym3    | 0.2747  | **0.2731** | 0.3652 | **0.3123** | **0.3877** | 0.3396 | **0.3363** | 0.3946 |
| Coif1   | 0.2609  | 0.2350  | 0.3807  | 0.2290  | 0.2209  | **0.3419** | 0.2486 | 0.3313 |
| Bior1.3 | 0.2626  | 0.2054  | **0.4710** | 0.2567 | 0.3417 | 0.2646 | 0.2047 | 0.4148 |
| Rbio1.3 | 0.2168  | 0.2711  | 0.2368  | 0.2488  | 0.1713  | 0.2498  | 0.2053  | **0.4157** |
| Dmey    | 0.1777  | 0.2548  | 0.3116  | 0.1277  | 0.1869  | 0.2781  | 0.1796  | 0.2429  |

Table 4. Values of mean global complexity for corresponding plots

|         | Fig. 13 | Fig. 14 | Fig. 15 | Fig. 16 | Fig. 17 | Fig. 18 | Fig. 19 | Fig. 20 |
|---------|---------|---------|---------|---------|---------|---------|---------|---------|
| Haar    | 0.2197  | 0.2395  | 0.3526  | 0.2746  | 0.2848  | 0.2827  | 0.2177  | **0.3954** |
| Db2     | 0.2960  | 0.3247  | 0.3249  | 0.2933  | 0.3732  | 0.3120  | 0.3245  | 0.3693  |
| Sym3    | **0.3020** | **0.3299** | **0.3975** | **0.3391** | **0.4482** | 0.3416 | **0.4036** | 0.3925 |
| Coif1   | 0.2684  | 0.2955  | 0.3520  | 0.2654  | 0.2933  | **0.3619** | 0.3004 | 0.3592 |
| Bior1.3 | 0.2331  | 0.2113  | 0.3625  | 0.2620  | 0.2918  | 0.2515  | 0.2133  | 0.3888  |
| Rbio1.3 | 0.2173  | 0.2531  | 0.3384  | 0.2470  | 0.2714  | 0.2788  | 0.2377  | 0.3560  |
| Dmey    | 0.1835  | 0.3021  | 0.3473  | 0.1643  | 0.2659  | 0.3306  | 0.2331  | 0.2843  |

## 6. Conclusion

In this paper, fractality is comprehended as the emergence of the same figure at different scales defining specific fractal geometry of icon referred to as personalized geometry. Considering the Orthodox icons of Byzantine style in these terms, two types of fractality are discerned – the descending and the ascending, where the former one corresponds to content of the icon and the latter one to self-organizing spatiality with the reverse perspective scaling which establishes the time axis. Overcoming of causal necessity manifested through the complexity increase in the temporal domain reveals the observer's personality scaling along the time axis as the origin of the fractal geometry of icon. In this sense, remarked fractal patterns, such as the von Koch`s curve and the fractal tree, become an expression of the personalized geometry that is not limited to the icon contents, but assumes the observer as the principal personality of the apparent world in which the icon also takes place objectively. The model has been confirmed as an adequate conceptual framework disclosing self-organization through the form of personalized fractality and yielding results that are completion of the visual observation of the icon.

**Acknowledgement:** The authors acknowledge support by the Ministry of Education, Science and Technological Development of the Republic of Serbia through the projects OI 174014, OI 179048 and III 44006.